\documentclass[useAMS,usenatbib]{mn2e}
\usepackage{graphicx}
\usepackage{amssymb}

\title[An Improved Model of SiO Maser Emission in Miras]
      {SiO Maser Emission in Miras}

\author[M.\,D.\ Gray, M.\ Wittkowski et al.]
       {M.\,D.\ Gray$^{1}$, M.\ Wittkowski$^{2}$, M.\ Scholz$^{3}$,
        E.\,M.\,L.\ Humphreys$^{4}$, K.\ Ohnaka$^{5}$, D.\ Boboltz$^{6}$\\ 
        $^{1}$ Jodrell Bank Centre for Astrophysics, Alan Turing
	       Building, University of Manchester, M13 9PL, UK\\
        $^{2}$ ESO, Karl-Scwarzschild-Str. 2, 
               85748 Garching bei M\"{u}nchen, Germany\\
        $^{3}$ Zentrum f\"ur Astronomie der Universit\"at Heidelberg (ZAH),
               Institut f\"{u}r \\Theoretische Astrophysik,
               Albert Ueberle-Str. 2, 69120 Heidelberg,
               Germany,\\and Institute of Astronomy, School of
	       Physics, University of Sydney, Sydney NSW 2006, Australia\\
        $^{4}$ Harvard-Smithsonian Center for Astrophysics,
               Cambridge, MA, USA\\
        $^{5}$ Max-Planck-Insitiut f\"{u}r Radioastronomie,
               Auf dem H\"{u}gel 69, 53121 Bonn, Germany\\
        $^{6}$ U.S. Naval Observatory, 3450 Massachusetts Avenue,
               Washington DC 20392-5420, USA}

\date{Accepted ... .
      Received ... ;
      in original form ...}

\pagerange{000--000}

\begin{document}

\maketitle

\label{firstpage}

\begin{abstract}
We describe a combined dynamic atmosphere and maser propagation
model of SiO maser emission in Mira variables. This model rectifies
many of the defects of an earlier model of this type, particularly
in relation to the infra-red (IR) radiation field generated by
dust and various wavelength-dependent, optically thick layers.
Modelled masers form in rings with radii consistent with those
found in VLBI observations and with earlier models. This agreement
requires the adoption of a radio photosphere of radius approximately
twice that of the stellar photosphere, in
agreement with observations. A radio photosphere of this
size renders invisible certain maser sites with high
amplification at low radii, and conceals high-velocity shocks,
which are absent in radio continuum observations. 
The SiO masers are brightest at an optical phase of 0.1 to 0.25,
which is consistent with observed phase-lags. Dust can have both
mild and profound effects on the maser emission. Maser rings,
a shock and the optically thick layer in the SiO pumping band
at 8.13\micron \,appear to be closely associated in three out of
four phase samples.
\end{abstract}


\section{Introduction}


We present preliminary results from a significantly improved model of
SiO
maser emission from the circumstellar envelopes of Mira-type, and
similar, long-period variable stars. The predecessor of this model,
hereafter the `old model',
was used to compute synthetic spectra and images of SiO maser emission
from a model star with parameters based on those of o~Cet
 \citep{ghf95,liz96}. With additional phase
information, the variation of the spectra and maps was computed over
the stellar period of 332\,d \citep{liz97,gh00,liz02}. 
The hydrodynamic solutions which
provided the input physical conditions for the maser sites in the old
model were taken from envelope computations by \citet{bowen88},
except in the earliest of these papers, \cite{ghf95}, which used
similar calculations by \citet{willson87}.

The old model used the (radial) density, velocity, and gas kinetic
temperature profiles from hydrodynamic
solutions by \citet{bowen88} and
 \citet{willson87}. These hydrodynamic solutions also provided
a radiative equilibrium temperature, $T_{eq}$, on the basis of
geometrical dilution of the stellar continuum, and on the opacity of
the gas, but no estimate of the abundance of SiO.
The radius of the stellar photosphere was recalculated
at each time step, and defined as the radius where the 
optical depth equalled 2/3, but the effective temperature of the
photospheric layer, whatever its radius,
had a constant value, independent of the stellar
phase. Dust was included in the pulsation model,
and was dynamically significant at radii beyond the condensation
zone \citep{bowen88}. At radii where dust was present, the
dust temperature was set equal to $T_{eq}$. The effect of dust
was exerted through a spectrally averaged
radiation pressure cross section,
a function of $T_{eq}$ and the dust condensation temperature.
The solutions did not specify a wavelength dependence for the
dust opacity, or a dust number density, though this latter quantity
could have been calculated, using sensible assumptions for the
dust mass fraction, grain radius and mineral density
(see \citealt{bowen88}). Later Bowen models improved upon
the version discussed above in several respects
\citep{bowen90,willson00}, but these later models were never used by us
for maser calculations.

The old model made several simplifications and
approximations, not inherent in the Willson and Bowen solutions,
for constructing the IR radiation field for
maser pumping. The stellar component of IR radiation field
assumed a fixed-radius stellar photosphere, and was
modelled as a black-body at the
(constant) stellar temperature of 3002\,K. Optically thin dust emission
was crudely modelled as a black-body at a single temperature,
weighted by a power-law in wavelength \citep{liz96}.
As a consequence of these
approximations, the radiative
pump rate in any SiO rovibrational transition had no intrinsic
variation with phase. Although the collisional part of the maser
pump was modelled more accurately than the radiative part, a
radiative component to the maser pumping was, however, always present
\citep{liz96,liz02}. The old model used a constant value of
10$^{-4}$ for the abundance of SiO, independent of both radius
and phase, and, for the purposes of pumping calculations, hydrogen
was assumed to be in molecular form, as no suitable set of
rate coefficients was then available for the H+SiO system. 

The stellar phase in the Bowen hydrodynamic solutions
was defined as zero
at the time a new shock left the
stellar photosphere \citep{bowen88,liz96},
and it was never possible reliably to relate this model phase to
the optical, or near IR, light curves of the star, despite several
attempts to do so, for example \citet{gray98}, 
\citet{liz02}. It was therefore
impossible to answer questions related to observed links
between the light-curves of SiO masers and the optical and
near-IR continuum, for example \citet{pardo04}. Polarization
and the presence of minor isotopomers of SiO were also not
included in the old model; these aspects of the problem remain
absent in the new model, and will not be considered further
in this work.

Despite the drawbacks listed above, the old model was surprisingly
successful when compared  with observations. The single phase
version \citep{ghf95,liz96} predicted new masers with upper rotational
states above $J$=6. Such masers were subsequently detected in a number
of stars, including examples from Miras and semi-regular
variables \citep{gea95,lizea97,gea99}.
 The ring structure of the SiO masers, found frequently in VLBI
observations \citep{philetal94,linc95,boboltz97}, 
which indicates predominantly
tangential amplification from a restricted range of radii, was
also strongly supportive of the model. It is worth noting here
that the model generated potential maser sites randomly throughout
a sphere of significantly larger radius than the ring in which
masers appear. If the model had been seriously inappropriate, it
could have produced, for example, a smooth radial distribution, or,
for predominantly radial amplification, a concentration of bright
maser spots directly over the central star. The ring radius
for the $v=2, J=1-0$ masers was found to be smaller, in a number
of stars, than for $v=1, J=1-0$, in agreement with the model
\citep{desmurs99,desmurs00,yi00,soria04,bob05}.
The model prediction for the ring radius of the masers in the
$v=1, J=2-1$ line at 86\,GHz, is that it should have a comparable,
or slightly larger radius than the ring formed by the $v=1, J=1-0$
masers, and be significantly less well populated. The second
prediction seems to be in ageement with observations 
\citep{soria04} but not the first. Observations of R~Leo
\citep{soria07} confirm that, in this star at least, the $v=1, J=2-1$
transition forms in a ring further from the star than the $J=1-0$
transition from the same vibrational state.
Single-dish observations of a number of Miras with different dust
temperatures \citep{naka07} showed a correlation of the
flux in the 8\,$\mu$m continuum and the $J=1-0$ masers from $v=1$ and $2$,
which strongly suggests a radiative component to the maser pump. 

The maser line ratios appear to be acceptable, given the
uncertainties in what constitutes the background for
unsaturated masers, except for the 
$v=2, J=2-1$ line which is much too strong when compared against
observations. However, the weakness of this maser in real sources
is very probably due to a line overlap with a water transition,
for example \citet{soria04},
and a molecular model of water was (and is) not included in our
maser model.

When phase dependence is added, the old model remains broadly
in agreement with interferometric observations, though with
some discrepancies. For example the model prediction of smaller
maser rings, at $v=2, J=1-0$, compared with
$v=1, J=1-0$, over the stellar cycle, is in agreement with
observations \citep{cotton04,yi05,cotton06,markus07}, as is
 the prediction
of an r.m.s. amplitude variation of 5-10 per cent in the
maser ring radius. 
These maser rings are also found
to be inside the inner radius of the silicate dust shell, but 
Al$_{2}$O$_{3}$ dust may be co-located with the masers \citep{markus07}.
However, the predicted relation between
SiO maser ring radius and SiO maser luminosity is not well
reproduced, and varies both between different stars, and
for different pulsation cycles of the same star \citep{cotton04,
cotton06}. Long term monitoring of the $v=1, J=1-0$ line in the
Mira TX~Cam \citep{phil03} shows motions dominated by shock-driven
outflow for part of the stellar period, whilst the remainder is
dominated by infall under gravity. This behaviour agrees with the
model assuming a reasonable link between the model and optical
phases, though the variation in the maser ring radius is
somewhat larger (18 per cent) in TX~Cam.
It should also be noted that TX~Cam exhibits considerable
asymmetric behaviour, and some maser features move outwards, for
example, even during the infall dominated phase.
If the model to optical phase relation used
by \citet{yi05} is adopted (model phase = 0.78 at optical maximum)
then the disrupted, weak maser ring structure found near
optical minimum in TX~Cam is also consistent with observations,
though any linking of the old model and optical phases should be
treated with great caution for reasons discussed above.

Single
dish monitoring observations suggest that shock waves, at least
of a type which take a significant fraction
of the period to cross the gap between
the stellar photosphere and the maser zone,
cannot be responsible for the maser pumping, though
some observations still suggest a coupling between the pump and
the kinematics of the envelope gas. See \citet{cotton06} for a
detailed discussion of the problems involved in relating results
from interferometric and single-dish observations.
Whilst it is not true to say that the old model is entirely
collisionally pumped, a major drawback is that it cannot explain
correlation between the light curves of SiO masers and radiation
in the IR pumping bands. It is clear that such correlations exist:
\citet{pardo04} found that the optical, near-IR continuum and
SiO masers share the same period in a sample of 13 Mira variables.
They also found the IR and SiO maser peak luminosities coincided
to within 0.05 of a stellar period. The IR data in
\citet{pardo04} is at 4.9 and 3.5\,$\mu$m, and therefore does not
include the crucial $\Delta v = 1$ pumping band of SiO at
8\,$\mu$m. \citet{mac06a} finds that the maxima in the 43\,GHz
masers from $v=0,1,2$ occur simultaneously to within $0.066$ periods
in o~Cet. As the $v=0$ maser ring was apparently much larger than
those at $v=1$ and $2$, this suggests that the pumping is predominantly
radiative. However, the old model makes no prediction about $v=0$
masers, and the difference between the ring radii at $v=1$ and $2$ is
sufficiently small ($\sim$0.1 to $\sim$0.2\,AU, dependent on phase)
 that a shock of reasonable speed could propagate
between them in 0.066 of a period (22\,d). These observations
\citep{mac06a} also did not detect any disruption of the maser
activity, due to shock arrival,  near optical minimum. Further
analysis with the same dataset from o~Cet \citep{mac06b} finds that
SiO maser flux density in the $v=1, J=1-0$ transition is correlated
with the velocity centroid of the emission. \citet{mac06b} offers
two possibilities for this correlation: either the pump affects
the kinematics of the maser region (that is, shocks strongly
influence maser amplification, principally through disruption of
velocity-coherent gain paths, but also by heating and compressing
the gas)
or that the maser pump and (independent?)
kinematic effects combine to produce the observed maser spectra.
The current authors find this second possibility hard to reconcile
with the model, since in the large velocity gradient 
(LVG) approximation used, the Sobolev
optical depth in any direction is dependent upon the velocity
gradient,
and therefore the kinematics are intimately coupled to the
pumping scheme. Similar observations for the Mira R~Cas \citep{mac07}
largely agree with those from o~Cet. However, it is interesting
that for one maximum, the peaks of the $v=1$ and $2$, $J=1-0$ masers
are significantly separated in phase.

The pulsating atmospheres of Miras are very extended because of
dynamic events, such as shocks. These extended atmospheres
exhibit phase-dependent temperature and density stratifications,
optically thick molecular layers, and dust formation; see for
example \citet{scholz03}. Observed radii of Miras have been found
to depend on both phase and wavelength of observation, which
is consistent with the predictions of dynamic models, in which
molecular layers lie above the continuum-forming layer
\citep{thompson02,perrin04,ohnaka04,ohnaka06b,markus08}.
In a series of experiments \citep{bob05,
markus07} mid-IR
interferometric measurements made with the VLTI were made
near contemporaneously with VLBA observations of the 43\,GHz
masers in $v=1$ and $2$ towards the Mira S~Ori. Models
\citep{ireland04a,ireland04b} were fitted to four observing
epochs, with an estimated error of $\lesssim$0.15 pulsation periods between
the model and optical phases. Both the IR photosphere radius,
defined as the radius at 1.04\,$\mu$m, and the radius of the
optically thick layer near 8\,$\mu$m (the $\Delta v = 1$ SiO
pumping band) vary in phase with the optical light curve, as far
as can be judged from the limited phase sampling. Importantly, the
radius of the 43\,GHz maser shells is slightly larger than this
8\,$\mu$m quasi-photospheric
band radius at all four observing epochs. If this
optically thick layer represents high-density post-shock gas,
then the position of the shock would be tightly correlated with
the light curve at this frequency, very possibly leading to a
resolution of the radiative/collisional pumping debate.
Recent work by \citet{woodruff08} suggests that, with certain
caveats, the conclusions obtained for S~Ori can be extended to
a sample of other Mira variables.

\section{Improvements to the Model}
\label{improve}

Our improved model removes many of the defects inherent in the old
version because of its restricted information about conditions in the
circumstellar envelope. In particular, the new version includes
radii for several optically thick layers, for both continua
and restricted-wavelength bands, as a function of phase.
The set of bands includes
all the IR pumping
bands of SiO which are accessible to the model. Other
features of the new model include
an accurate estimate of the IR radiation field from
dust, sufficient chemistry to provide abundances for SiO and its
main collision partners, and new rate coefficients for collisions of
SiO with hydrogen atoms. A link between the model phase (in terms of
the shock position) and the stellar light curve is implicitly
included in the model data.

At this point we digress to make some definitions regarding the
optically thick layers introduced above. We reserve the word
`photosphere' to mean a continuum-forming layer, and we always
use this word with a modifier which specifies the region of the
spectrum to which it applies. In particular, we define `IR photosphere'
to be the continuum-forming layer at 1.04\,\micron \,where the
optical depth is equal to 1. This wavelength was chosen because
it is in a region little contaminated by molecular line absorption. 
We use `radio photosphere' to be the
continuum-forming layer in the radio region. This is less well
defined, but, in the context of SiO, can be taken to mean the region
of the electromagnetic spectrum occupied by the pure rotational
lines of SiO in general, and the maser lines at
43 and 86\,GHz in particular. When referring to other work, 
for example the hydrodynamic solutions of the old model, which does not
match the precise definition of the IR photosphere, we use the
expression, `stellar photosphere', to describe the continuum-forming
layer in the optical and near IR. Where photospheric radii have already
been mentioned in the Introduction, they conform to the above
definitions.

Each SiO pumping band has a quasi-photospheric, optically thick,
layer, due to the presence of large numbers of molecular lines,
mostly from water. These layers have radii computed as
$\tau_{\lambda}=1$ radii based on the definition
of the filter radius in \citet{scholz87} using a box filter
of width 0.1\,\micron. When referring to these optically
thick layers, the form of words we adopt is 
`optically thick band radius', or
`optically thick layer'. When referring only to the
radii of such layers, we write a shorter form, specifying
the wavelength of the band in microns, followed by `radius'.
For example, to refer to the radius of the optically thick
molecular layer in the $\Delta v = 1$ band of SiO, we use the
shorthand `8.14\,\micron \,radius'.

\subsection{Hydrodynamic Solutions}
\label{improve_hydro}

The hydrodynamic solutions used in the present work are based on
those published in \citet{ireland04a} and \citet{ireland04b}. We
are particularly interested in the M-series of models, drawn from
the latter work. All these models have a pulsation period of
332\,d, like the real star o~Cet and the old model. However, other
parameters of the M-series models might be more appropriate for a
model of R~Leo.
The abundance of SiO is computed by an additional
code by Ohnaka. This code
 employs equilibrium chemistry, including 34 elements and
235 molecular species, and
assumes solar abundances of the reactants.
Optically thick band radii are provided in the SiO 
rovibrational pumping bands at 2.03, 2.74, 4.09 and 8.14\,$\mu$m,
corresponding to vibrational changes, respectively, of
$\Delta v =$ 4, 3, 2 and 1.

\begin{table}
\caption{Parameters for the M-Series models \citep{ireland04b}.
The radius R$_{p}$ refers to the parent star (see text). The
effective temperature and luminosity are computed for the
non-pulsating parent star.
}
\label{sparms}
\begin{tabular}{@{}llr}
\hline
Parameter    &  Symbol        & Value    \\
\hline
Period       &   P            & 332\,d   \\
Mass         & M/M$_{\sun}$   & 1.2      \\
Metallicity  & z/z$_{\sun}$   & 1.0      \\
Luminosity   & L/L$_{\sun}$   & 3470     \\
Radius       & R$_{p}$/R$_{\sun}$ & 260    \\
Temperature  & T$_{eff}$       & 2750\,K  \\
\hline
\end{tabular}
\end{table}

The M-series of models is decribed in detail in \citet{ireland04b}.
The important stellar parameters are briefly listed here
in Table~\ref{sparms}. The radius and effective temperature
refer to a non-pulsating `parent' star, where these values are
independent of phase. The parent parameters are based on a layer
where the Rosseland mean optical depth is 1.
There are 20 M-series models altogether, some with the subscript n,
denoting a new model, first published in \citet{ireland04b}. Other
models appeared previously in either \citet{hsw98} or \citet{tlsw03}.
 Each model has a cycle (pulsation) number
and a phase within that cycle. At present the samples range from
phase 0.49 of cycle 0 to phase 0.50 of cycle 2, giving
ten models per cycle, though these are not evenly spaced in phase.
Model parameters are listed in Table~1 of \citet{ireland04b}. Phase
values are given in terms of the optical light-curve and are
estimated accurate to about 0.1 pulsation periods absolutely, but more
accurately relatively (0.01-0.02 periods). An estimate
of the phase-variable 
IR photospheric radius is given as
$R_{1.04}$ for each model. The basic models are dust free, and
extend radially to 5\,R$_{p}$, where R$_{p}$ is the radius of the
parent star. This is a considerably smaller radial coverage than
in \citet{bowen88}, but this has no consequences if the water and
OH maser zones are ignored. We also note that the models we use are
developed from grey dynamic models which extend to considerably
larger radii than  5\,R$_{p}$, and which establish the initial
pressure stratification for the refined 
non-grey models.

In this work, we use the same models which were used in the
fit to S~Ori \citep{markus07}. These are M21n, M22, M23n and
M24n, which have respective optical phases of 0.10, 0.25,
0.30 and 0.4. The stellar parameters of S~Ori differ from those used
for the M~series of models in pulsation period, radius and mass. The
M-models were used as the best available option to describe the
dynamic atmosphere of S~Ori, scaling the period to phases between
0 and 1, and the radius to match the observed angular diameter
of S~Ori. See also the discussion in \citet{markus07}. We do
not expect these scaling operations to lead to strong effects, but
it should be kept in mind that remaining differences between model
predictions and observations of S~Ori might arise from this moderate
mis-match of stellar parameters. 

\begin{figure}
\includegraphics[width=84mm]{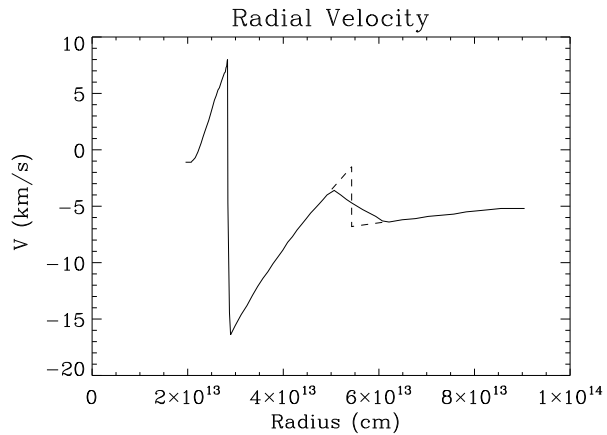}
\caption{Radial velocity of model M21n as a function
of radius, showing the presence of outward-moving
shocks. The shape of the outer shock appears smoothed in the
model for computational reasons (dashed line: without smoothing
applied).
}
\label{newv}
\end{figure}

\begin{figure}
\includegraphics[width=84mm]{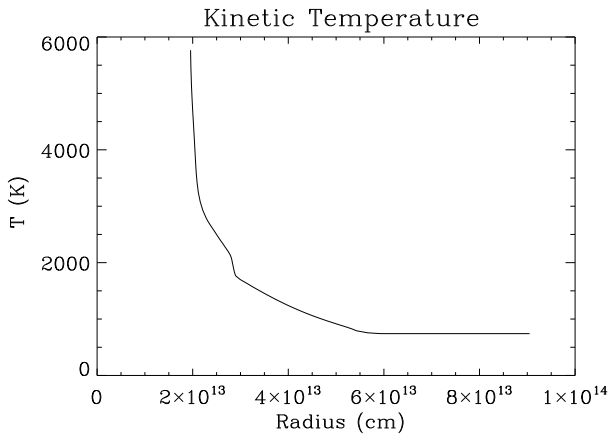}
\caption{Kinetic temperature of the gas in model M21n:
the constant temperature of 741\,K at large radii is
artificially set, since the equation of state is poorly
known at the very low temperatures of the uppermost layers.
}
\label{newtk}
\end{figure}

\begin{figure}
\includegraphics[width=84mm]{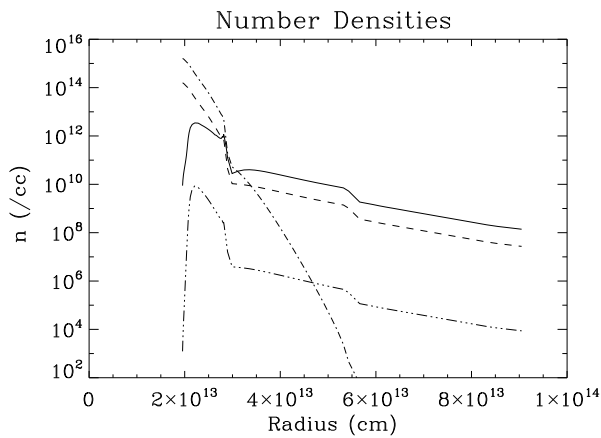}
\caption{Number densities of H$_{2}$ (solid line),
He (dashed line), H atoms (chained line) and
SiO (complex line)
}
\label{newdn}
\end{figure}

As examples of the behaviour of the physical
conditions in the model atmosphere
we plot, for M21n, the radial velocity (positive means
outflow), kinetic temperature, and the number densities of
SiO and its main collision partners as functions of radius in
Figures~\ref{newv}, \ref{newtk} and \ref{newdn}, respectively.
A strong inner, and
a weaker outer, shock are visible in Figure~\ref{newv} at
radii of 1.57 and 3.00\,R$_{p}$
(2.84$\times$10$^{13}$ and 5.43$\times$10$^{13}$\,cm). 
The shock fronts are slightly
smeared out in the models \citep[see][]{ireland04b} and affect
both the temperature and density graphs.  
The shape
of the SiO number density curve follows closely that of
H$_{2}$. The optically-thick band radii in the SiO
pumping bands at 2.03, 2.71, 4.06 and 8.13\,$\mu$m are,
respectively, (2.51,4.18,2.13 and 3.98)$\times$10$^{13}$\,cm,
so the 8.13\,$\mu$m layer, at this phase, actually lies
outside the radius of the strong, inner shock.

\subsection{Dust and the IR Radiation Field}
\label{dustandir}

In \citet{markus07}, the dust-free hydrodynamic solutions 
described above were complemented by ad-hoc radiative transfer 
models of the dust shell, using the radiative transfer 
code {\sc mcsim\_mpi} by \citet{ohnaka06a}. A dust-free model 
of the M series was combined with a dust shell model using
dust shell parameters so that the overall models fitted 
the 4 epochs of observations of S Ori. 
The best-fitting overall models for the four epochs of observation,
A, B, C, D have parameters listed in Table~\ref{epochtab}.
\begin{table}
\caption{Standard pairings of dust epochs and hydrodynamic
solutions with parameters of the model star and dust shells.
Observational phases ($\phi_{obs}$, Column~2)
refer to S~Ori \citep{markus07}. Theoretical phases ($\phi$,
Column~4) are the phases of the hydroynamic model, on the
optical scale, 
fitted to each epoch. Column~5 is the ratio of the inner radius
of the dust shell to the IR photospheric radius, and
Column~6 is the visual optical depth of the dust shell. 
}
\label{epochtab}
\begin{tabular}{@{}lrrrrr}
\hline
Epoch &$\phi_{obs}$& Model  &$\phi$& $R_{in}/R_{1.04}$ &$\tau_{V}$\\
\hline
  A   & 0.42       &  M22   &  0.25& 1.8               & 2.5   \\
  B   & 0.55       &  M24n  &  0.40& 2.0               & 2.5   \\
  C   & 1.16       &  M23n  &  0.30& 2.2               & 1.5   \\
  D   & 1.27       &  M21n  &  0.10& 2.4               & 1.5   \\
\hline
\end{tabular}
\end{table}
We therefore adopt as standard, the pairings
epochA+M22, epochB+M24n, epochC+M23n and
epochD+M21n, and use these in the work that follows,
except in Section~\ref{varidust} and Section~\ref{resdust}.

In the modelling of S~Ori, combinations of Al$_{2}$O$_{3}$
and silicate dust were tested, but better fits were obtained
from dust-shell models without silicates. Therefore,
the epochA-D dust shells use Al$_{2}$O$_{3}$ dust
only, which is assumed to be in the form of spherical grains
of a single radius, a = 0.1\,$\mu$m. Each dust epoch provides,
as a function of radius, the number density and temperature
of the dust. Optical efficiencies for absorption and scattering
were taken from \citet{koike95} for wavelengths shortward of
7.8\,$\mu$m, and from \citet{begemann97} for longer wavelengths.
These optical efficiencies, the table of number
densities and temperatures, together with stellar parameters
provide all the information necessary
to compute the angle-averaged
 intensity of the IR dust radiation field
in the circumstellar envelope (see Section~\ref{computations}).

Silicate dust can optionally be included in the dust shell
calculations. When it is so included, the optical efficiencies
are the `warm silicate' variety
from \citet{ossenkopf92}. The silicate grains are assumed
to be spheres with the same single
 radius as their Al$_{2}$O$_{3}$ counterparts.
Dust of this type has been used in a
model of the envelope of the Mira GX~Mon
(Boboltz et al., in preparation). We note that the dust shell
extends to substantially larger radii 
(R$_{in}\sim$2\,R$_{p}$ and R$_{out}$/R$_{in}$=1000)
and, when present, the silicate dust is found at
larger radii (inner radius $\sim$4\,R$_{p}$) than the Al$_{2}$O$_{3}$ dust.

\begin{figure}
\includegraphics[width=84mm]{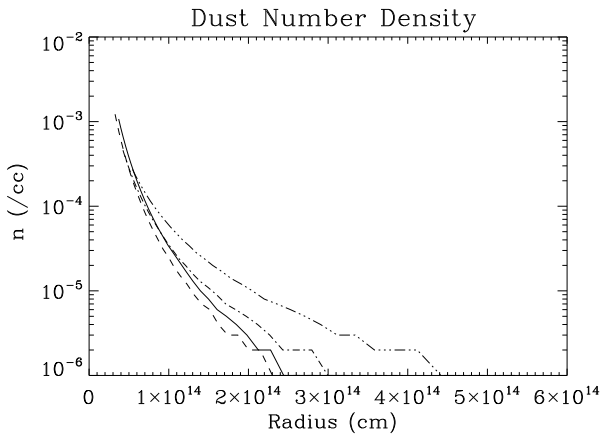}
\caption{Number density of Al$_{2}$O$_{3}$ dust as a function of
radius for the four epochs A (solid line), B (dashed line),
C (chained line), and D (complex line)
}
\label{newndust}
\end{figure}

\begin{figure}
\includegraphics[width=84mm]{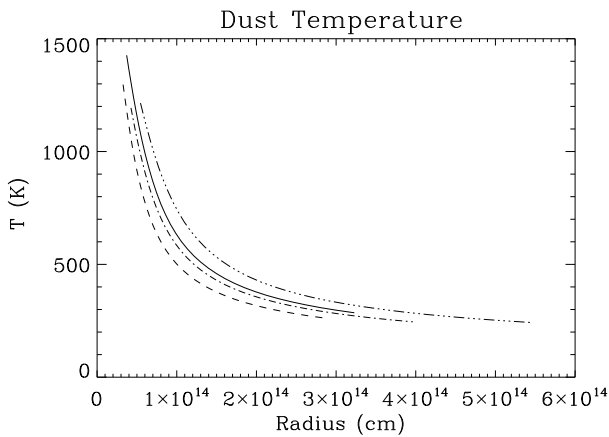}
\caption{As for Fig.~\ref{newndust}, but showing the
dust temperature. Epoch~B has the smallest inner radius,
followed by Epoch~A. Epoch~A has the highest temperature
at its inner radius.
}
\label{newtdust}
\end{figure}

We plot, in Fig.~\ref{newndust}, the number density of Al$_{2}$O$_{3}$ dust as
a function of radii for the four epochs A-D. Fig.~\ref{newtdust} shows
the dust temperature as a function of radius for the same four
models.

As noted in the introduction, dust was dynamically important in the
hydrodynamic solutions of the old model, but not so with the dust
shells attached to the dust-free hydrodynamic solutions of the current
model. In this connexion, we note that studies by 
\citet{ireland06} found that dust did not noticeably affect the
density
and temperature stratification of the model for radii
$\lesssim$ 2\,R$_{p}$ in Miras with parameters similar to o~Cet or
R~Leo. Preliminary results from very recent calculations
\citep{ireland08}, which do include the formation and
dynamical effects of dust, suggest that this view is correct,
though only a small subset of the full phase and cycle data
is currently available. These conclusions from models are supported
by measurements of the inner dust 
radius with MIDI \citep{markus07}, which
place it in range 1.8-2.2 IR photospheric radii.

\subsection{Rate Coefficients for H+SiO}

For details of the rate coefficients used for collisions
between H$_{2}$ and SiO, see \citet{liz96} and references
therein. These rate-coefficients
were originally computed for He+SiO collisions, and can easily
be reset to the latter system by changing the total and
reduced masses.

The inner regions of the circumstellar envelope contain 
substantial amounts of hydrogen 
in atomic form (see Fig.~\ref{newdn}). We have therefore
included a set of rate coefficients for H+SiO collisions. For
details of the computations of the 
potential surface and collision cross sections,
see \citet{jimeno99} and \citet{palov02}; for calculation
of the rate coefficients from
these cross-sections see \citet{palov06}. Rate coefficients
can be computed for any rovibrational transition in the
first six vibrational states ($v=0-5$) for rotational states
from $J=0-40$, and
with a maximum 
change in rotational quantum number of $|\Delta J|=40$.
The accuracy of the rate coefficients degrades progressively
above 5500\,K, but this is not a problem with the hydrodynamic
solutions used in the present work, where the maximum kinetic 
temperture is $\sim$6000\,K (see Fig.~\ref{newtk}). The
H+SiO rate coefficients are different, and 
systematically larger than for
the He+SiO system. They are of similar magnitude for
$\Delta v=0$ transitions, but larger by a factor of 10 to
100 for the vibrationally inelastic case. The expected
decay in the magnitude of rate coefficients as $\Delta v$
increases is also rather slow, so that H+SiO collisions can
efficiently make rather large changes in vibrational state.

\section{Remaining Problems}
\label{problems}

The improvements discussed in 
Section~\ref{improve} do not solve all the
problems with the model, and actually introduce one or two
new, though minor, difficulties. Most of these relate to
our lack of knowledege of the position of the radio photosphere.
In addition, we still rely on the
LVG approximation for the maser pumping code, and we have
no rate coefficients which are based on the true
H$_{2}$+SiO collisional system, though some new calculations
may soon become available \citep{bienieck08}.

\subsection{The radio photosphere}
\label{prob_photo}

The new hydrodynamic solutions
 (Section~\ref{improve_hydro}) provide phase-dependent
radii for the optically thick
molecular layers in the four rovibrational pumping
bands which are possible in our model, which has 200
levels distributed equally between the first five vibrational
states ($v=0-4$). However, the pure rotational transitions
cannot be assigned to any of these bands, and the position
of the optically thick layer, or radio photosphere,
at the longer
wavelengths (sub-mm to cm) of these rotational lines is much less 
certain. The radiation field from the dust can be ignored
for the pure rotational lines because of the very low optical
efficiencies at these wavelengths, but see Section~\ref{prob_back}
below. Therefore, we need to consider only the stellar 
radiation from the point of view of pure 
rotational radiative maser pumping.

Estimates of the radius of the radio photosphere have
been made by \citet{reidmen97}, using three frequencies, all
lower than any SiO pure rotational lines. The opacity for the
radio photosphere is
maintained by free-free interactions of electrons with neutral
H and H$_{2}$, the electrons originating from ionization of K
and Na. Figure~10 of \citet{reidmen97} strongly suggests that the 
radius of the radio photosphere
at the SiO frequencies (in the approximate range
40-1500\,GHz) is smaller than at the measured frequencies, and
may change by a factor of $\sim$1.3 over this range of 
frequencies. Overall,
\citet{reidmen97} give a figure of 2\,R$_{*}$ for
the radius of the radio photosphere, where R$_{*}$ is the
radius of the stellar photosphere. The figure
of 2\,R$_{*}$ still appears in more recent work
\citep{reidmen07}, but the radio photosphere is also linked
to the optically thick molecular layer which becomes opaque 
in the mid infra-red.
With this limited information, it is difficult to make a
good choice for the radius of the radio photosphere, but we
have chosen to associate it with the molecular layer, as
in \citet{reidmen07}, and
we have set the radius of the radio photosphere
equal to the optically thick
band radius at the longest IR waveband (the 8\,$\mu$m radius).
This means that the radio photosphere, like the optically thick layers
in the IR bands, moves with the stellar pulsations. We also
note that if we can set $R_{*}\sim R_{p}$, then this choice
of radius in the radio region is in good agreement with the
figure of 2\,R$_{*}$. However, given the uncertainties in
the position of the radio photosphere, and the fact that
the maser lines cover an order of magnitude in wavelength,
so there is unlikely to be one optically thick radius for
all these lines, we allow the choice of maser sites to vary
from the smallest radius in the hydrodynamic solution, at each
phase, up to the constant value of 5\,R$_{p}$. We can then
investigate the effect of making the radius of the radio photosphere 
very small, and subject the site distribution to
later exclusions on the basis of the larger 8\,$\mu$m
radius discussed above.

\subsection{Maser backgrounds}
\label{prob_back}

In the old model, the background amplified by the masers
was set to a Planck function at the gas kinetic temperature
local to each maser site, without much justification for this
choice. However, given that the allowed masers in the model
are a subset of the pure rotational transitions discussed
above in Section~\ref{prob_photo} (those with upper $J\leq 10$),
the presence of a radio photosphere at radii similar to those
of the maser zone makes this choice rather more acceptable,
and we keep it for the new model. The alternative would be
to integrate along the line-of-sight through the dust shell
over all the dust and SiO spontaneous emission lying behind
the maser site. However, this background is optically very
thin, and would produce a far weaker background than optically
thick emission from the radio photosphere. Of course, for
 saturating masers,
it does not matter what the background radiation is. However,
the amplifying column required to reach saturation is 
significantly longer for a weak background, so that a
background originating from the radio photosphere
would, on average, lead to a brighter set of masers.

An interesting possibility is that it is the background which
breaks the symmetry of the maser shells in real stars, leading
to the distribution of spots that we see, instead of an
unbroken ring. In the old model, we appealed to variations in
the SiO abundance to break the symmetry, but as we now have
this as a smooth function of radius (see Fig.~\ref{newdn}) it
is less acceptable to do this. A purely speculative idea at
present is that the maser sites lie above
regions which are like extensions to the
radio photosphere, with enhanced electron
abundance from above average ionization of the alkali
metals, and consequently high input backgrounds to overlying
masers. These sites would then have the advantage over
surrounding regions of needing shorter amplifying columns
to reach the same maser intensity. Asymmetries of this
type in the molecular layer, which forms the optically
thick band radius at 8\,\micron \,have already been found
\citep{weigelt96,ragland08}, although it is not clear
that these would extend, in wavelength, to the radio region.
 The choice of maser sites
in our computations remains one of random selection
\citep{liz96}.

\subsection{Pointlike Masers}
\label{prob_point}

The use of the LVG approximation for the transport of the
pumping radiation simplifies the radiation transfer problem,
but there is a considerable price to pay for this simplicity.
The approximation is essentially local, so that a region
consitituting the maser is effectively defined by the distance
(which can vary with direction) required for the radiation
lineshape to be shifted out of resonance with the response
lineshape of the molecules. When we pick the coordinates of
a maser site,
there is still an assumption that
the physical conditions within this, velocity bounded,
LVG zone are equal to those at the selected coordinates.

\subsection{LTE Chemistry}
\label{prob_chem}

The abundances of the atomic and molecular species used in the current model
are based on LTE chemistry. This amounts to an assumption that 
the network of chemical reactions comes to equilibrium on a timescale
which is short compared with important hydrodynamic timescales in
the envelope. It is likely that non-LTE chemistry is important is
establishing the time-varying populations of a number of species,
including CO,HCN,CS and SiO, as predicted in models
by \citet{cherchneff06}. Detection of all these molecules in a
sample of AGB stars of widely varying C/O abundance strongly
supports the model predictions \citep{decin08}. There is a brief
discussion of non-LTE effects pertaining to models of the type used
in the present work in \citet{ireland08}.

\section{Computations}
\label{computations}

\subsection{IR Pumping Radiation}

We briefly consider the conversion of the dust optical efficiencies
described in Section~\ref{dustandir} to the inputs required by
the maser model: namely effective grain areas and angle-averaged
intensities ($\bar{J}$) at the wavelength of every electric-dipole-allowed
transition in the SiO model. The areas were trivially obtained
from the optical efficiencies, since the grains are modelled as
having a unique radius.

To compute the angle-averaged intensities, we solve, for each
tabulated dust wavelength and radius
in the dust model, the radiative transfer equation,
\begin{eqnarray}
\frac{dI}{d\tau} + I & = &
\frac{B(T_{A})}{1+(\kappa_{S}/\kappa_{A}) + (\sigma /\kappa_{A})}
\nonumber \\ & + &
\frac{B(T_{S})}{1+(\kappa_{A}/\kappa_{S}) + (\sigma /\kappa_{S})}
+ \frac{\bar{J}}{1+\kappa /\sigma},
\label{rteq}
\end{eqnarray}
which is written for the most complicated situation considered,
with a two-component dust shell of Al$_{2}$O$_{3}$ (subscript A)
and silicate (subscript S). Absorption coefficients ($\kappa$)
and scattering coefficients ($\sigma$) are written without a
subscript when summing over both dust types, for example
$\sigma = \sigma_{A} + \sigma_{S}$. The optical depth is given
by the standard expression, \( d\tau = (\kappa + \sigma )ds \),
where $ds$ is the element of distance along a ray. With the
single-radius dust approximation, it is possible to write the
ratios that appear in eq.(\ref{rteq}) directly in terms of the
optical efficiencies for absorption, $Q_{abs}$, and scattering,
$Q_{sca}$, for example
\begin{equation}
\frac{\kappa_{S}}{\kappa_{A}} = \frac{n_{S}Q_{abs,S}}{n_{A}Q_{abs,A}},
\label{kaprat}
\end{equation}
where $n_{S}$ ($n_{A}$) is the number density of silicate (Al$_{2}$O$_{3}$)
dust.
Similar expressions can be used for the other ratios.
The thermal emission terms
contain Planck functions, $B(T)$, for the appropriate dust
temperature, and the scattering term contains the angle-averaged
intensity, $\bar{J}$, defined in the spherical geometry used here as
\begin{equation}
\bar{J} = \frac{1}{2} \int_{-1}^{1} I(\mu ) d\mu .
\label{meanint}
\end{equation}
We note that as eq.(\ref{rteq}) contains $\bar{J}$, the solution is
iterative, with specific intensities for each value of $\mu =
\cos \theta$ being computed with the most recent estimate of
$\bar{J}$, before a new value is found from eq.(\ref{meanint}).
The boundary conditions used in the solution of eq.(\ref{rteq})
were to use a Planck function at the kinetic temperature
of the appropriate SiO band radius
for values of $\mu$ that intersected such an optically thick layer.
For other values of $\mu$, a Planck function at the
CMB temperature was used.
A value of $\bar{J}$ was calculated for each radius in the hydrodynamic
solution, and for each dust wavelength.

\begin{figure}
\includegraphics[width=84mm]{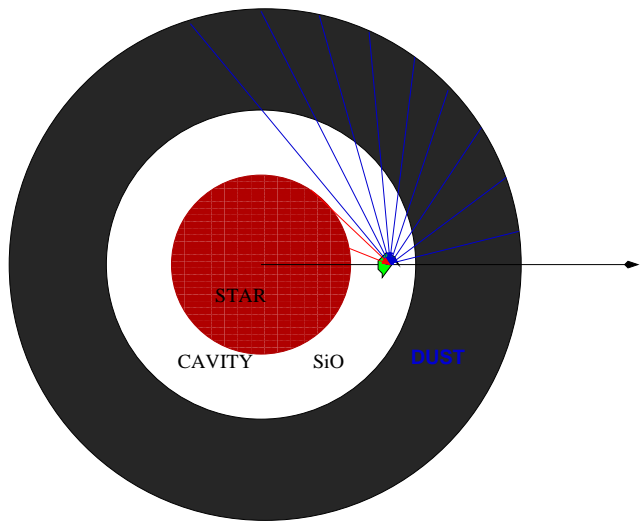}
\caption{A maser site shown in the cavity between an
optically thick layer, marked as the star,
and the surrounding dust shell (case (ii) in the text). Rays
at various possible angles are shown converging on the maser
site, where they are angle averaged. The size of the optically
thick layer is different, in general, for each SiO
pumping band and pulsational phase.
}
\label{dustcalc}
\end{figure}

We next calculate the dust contribution to the pumping rates
in each SiO transition. For a transition with upper level, $u$,
and lower level, $l$, the dust contributions are
$B_{ul}\bar{J}$ and $B_{lu}\bar{J}$, where $B_{ul}$ and
$B_{lu}$ are the Einstein B-coefficients for the transition.
We first assign
each transition to either one of the IR pumping bands, or as `radio',
a blanket designation for all the pure rotational lines. The
input data for the dust routine is then read in: the dust
areas tabulated by wavelength, the dust temperatures and
number densities tabulated by radius, and the dust mean intensities,
computed as discussed above, tabulated by both radius and
wavelength. These mean intensities are required for calculating
the scattering contribution to the dust radiation field.
Then, considering the radius of a single maser
site, a geometrical dilution factor is calculated relative
to the optically thick band radius at
each pumping band. In a loop over the
SiO spectral lines, we then classify the maser site as in
one of three basic positions for the current line. Case (i)
is that the site is inside
the band radius, in which case the pump in the
current line corresponds to
a black-body at the temperature of the site radius and wavelength
of the line; there is no dust contribution.
Case (ii) is that the maser site
is in a dust-free gap between the band radius and the
dust shell, for example Fig.~\ref{dustcalc}. The final possibility,
case (iii), is that the maser site is inside the dust shell.
In both case (ii) and case (iii), a
radiative transfer equation is solved for the current transition.
Rays were traced at a number of angles through the envelope, with
each ray terminating on the maser site. Values of the dust
areas, and mean intensities at the line wavelength were obtained
from spline interpolations from the input tables for each
site radius. A given ray had two possible starting conditions:
an optically thick layer, yielding a black-body at the 
temperature of that layer, or an interstellar value,
taken to be the CMB. Case (iii) required the stellar 
contribution to undergo some extinction due to the dust.
Angle-averaging of the rays produced the required mean intensity
at the wavelength of a particular line, and this was trivially
converted to a pump rate.

\subsection{Size of Radio Photosphere}

We have noted in Section~\ref{prob_photo} that there is
a considerable uncertainty in the size of the radio photosphere
for the maser lines. We here show the difference between a
model based on the standard value, which is equal to the
8\,$\mu$m radius, and an otherwise identical
model in which the radio photosphere has been taken to
coincide with the IR photosphere. 
In both cases we use Model~1
 (see Table~\ref{dusttab}), and the same distribution of
maser spots. In the first version of the model (1a) we allow
masers with nominal positions outside an IR photospheric radius of
1.21\,R$_{p}$ (the 1.04\,$\mu$m value) to grow; in Model~1b, we
exclude masers inside 2.20\,R$_{p}$, corresponding to the
8\,$\mu$m radius. This exclusion is based on the
idea that, though SiO molecules may exist at the site,
and these may be sucessfully pumped, optically thick layers
at larger radii, and at the maser frequency, will prevent
external observers from seeing the emission.

\subsection{Variation of dust envelopes}
\label{varidust}

Initially, we considered a single hydrodynamic solution, M21n, and
ran the maser code for 1500 maser sites, as in the old model,
for each dust epoch in turn. We note that M21n is actually
paired with dust epoch D, but we ran it with the other
three epochs simply to look for the effects of varying the
dust component of the radiative pump. The models testing the
variation of the dust envelope are listed in
Table~\ref{dusttab}.
\begin{table}
\caption{Jobs run to investigate the effect of variation of
the dust shell only. All other parameters are for the model
with structure M21n \citep{ireland04b}.
}
\label{dusttab}
\begin{tabular}{@{}lrr}
\hline
Job Number  & Structure  & Dust Epoch \\
\hline
    1      &  M21n      &  D          \\
    2      &  M21n      &  C          \\
    3      &  M21n      &  B          \\
    4      &  M21n      &  A          \\
\hline
\end{tabular}
\end{table}
The seed for the random number generator was the same for all these
models. As they also have the same structure, which means the same
parent radius for the star, the maser positions were the same
for all the models. Results for the dust variation models appear
in Section~\ref{resdust}.

\subsection{Phase Samples}

In this set of computations we ran jobs for four hydrodynamic
solutions, each with its standard dust epoch. Details
of these jobs appear in Table~\ref{phasetab}. Note that job 1
appears in both the dust variation and phase sample data.
\begin{table}
\caption{Jobs run to investigate the effect of stellar
phase on the intensity and distribution of maser spots.
Structure designations are from
\citet{ireland04b}. Phases are based on the
optical light curve. Stellar radii (Column~5) are based on the IR
photosphere,
at 1.04\,\micron , and given in units of the `parent' radius
of the star, where $R_{p}=1.809\times 10^{13}$\,cm.
8.13\,\micron \,radii,
assumed equal to the size of the radio photosphere, are
also in parent units.
}
\label{phasetab}
\begin{tabular}{@{}lrrrrr}
\hline
Job   & Structure  & Dust & Phase & $R_{1.04}/R_{p}$ &$R_{8.13}/R_{p}$\\
\hline
    1&  M21n      &  D   &  0.10 & 1.21 & 2.20   \\
    5&  M22       &  A   &  0.25 & 1.10 & 1.77   \\
    6&  M23n      &  C   &  0.30 & 1.03 & 1.73   \\
    7&  M24n      &  B   &  0.40 & 0.87 & 1.69   \\
\hline
\end{tabular}
\end{table}
The models all used 1500 maser sites, randomly distributed
between the (phase dependent) IR photospheric radius (see
Table~\ref{phasetab}) and an outer radius of 
5$R_{p}$. In view of the small number of phase samples,
we did not attempt to advance the model with phase, the
method adopted in \citet{liz02}, but instead we
re-randomized the maser positions for each new model.
Results appear in Section~\ref{resphase}.

\section{Results}

\subsection{Size of radio photosphere}

\begin{figure}
\includegraphics[width=84mm]{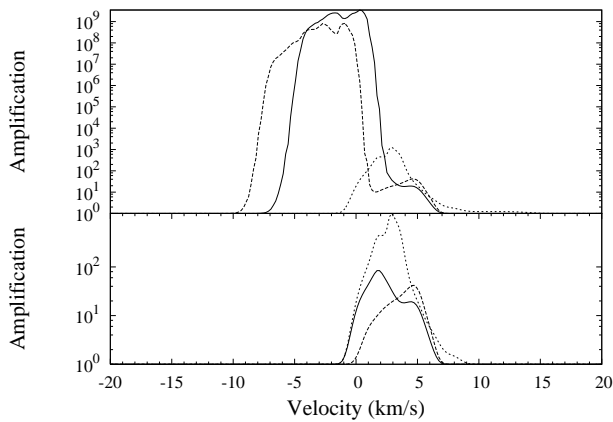}
\caption{Spectra of the three SiO maser lines $v=1, J=1-0$
(solid line), $v=1, J=2-1$ (dashed line) and $v=2, J=1-0$
(dotted line) for Model~1a (top) and Model~1b (bottom).
These model differ only in the assumed size of the radio photosphere,
which is smaller in Model~1a (upper graphs).
}
\label{spec_photos}
\end{figure}

In Fig.~\ref{spec_photos}, we show spectra from Model~1a (top)
and Model~1b (bottom). Recall that in Model~1a, the radio photosphere
is small, coinciding with the IR photosphere,
which at this phase has a
radius of 1.21\,R$_{p}$. By contrast, in Model~1b, the
radio photosphere coincides with the 8.13\,\micron \,radius,
equal to 2.20\,R$_{p}$.
The three spectra shown in each panel are the three most
commonly observed lines: $v=1, J=1-0$ and $J=2-1$ and
$v=2, J=1-0$. In both lines from $v=1$, Model~1a (top
panel) produces extremely high amplification masers.
These are emitted by a population of maser sites,
just outside the small radio photosphere of this model. 
We refer to this population as the `bright, inner object'
to distinguish them from more modest maser emitting
sites at larger radii in the envelope.

 The $v=2$ transition has the same shape in
both panels, so the bright, inner, objects do not emit
significantly in this line. Emission from the bright inner
masers in both the $v=1$ transitions can be seen to be
blue-shifted by $\sim$5\,km\,s$^{-1}$ relative to the
emission from the outer population. This shift arises
naturally if the inner population of masers is associated
with post-shock gas
inside the inner shock (see Fig.~\ref{newv})

\begin{figure}
\includegraphics[width=84mm]{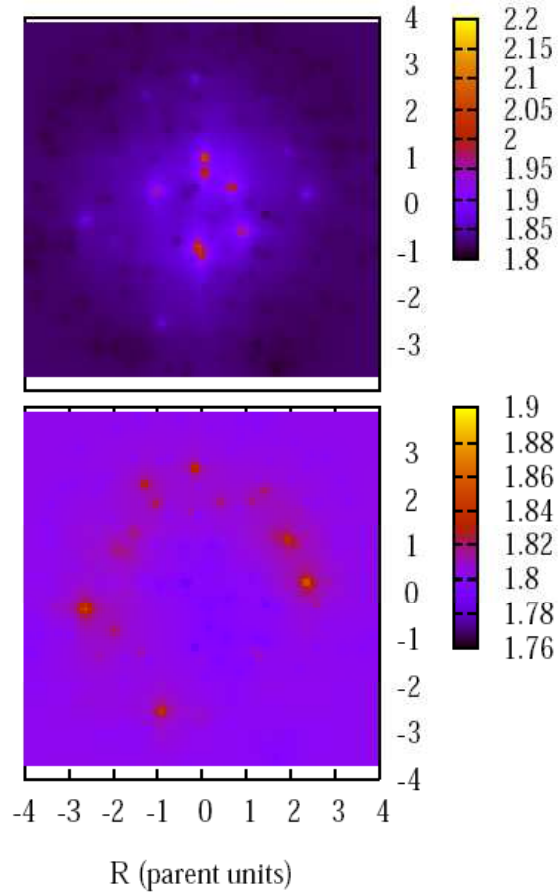}
\caption{Modelled maps of the $v=1, J=1-0$ transition
with radio photospheric radii of 1.21\,R$_{p}$ (top)
and 2.20\,R$_{p}$ (bottom). In the top panel, the log
to the base 10 of the
maser amplification
is clipped at a value of 3, for any
individual spot, to render the outer
ring visible.
}
\label{map_photos}
\end{figure}

In Fig.~\ref{map_photos} we show synthetic
maps of the 43\,GHz $v=1, J=1-0$ masers, the most commonly
mapped transition with VLBI instruments. In the upper figure,
we again have the smaller radio photosphere, allowing the
appearance of an inner ring of very bright masers, with
a projected radius close to $\sim$1\,R$_{p}$. The lower panel
shows the maser distribution with the larger radio
photospheric radius of 2.20\,R$_{p}$. The maser ring in
the lower panel has properties which are similar to those
of a `quiet' phase in the old model, and a radius which
is in good agreement with observations.
 If the spots of the inner ring had similar dimensions and
beam angles to those at larger radii, they would be a factor
of 10$^{6}$ times brighter than those found in the old
model, or at larger radii in the current model. As masers
of this power are not observed, we conclude that such a
small radio photosphere is not realistic, and we adopt the
larger, standard, value from here on.
We note that, even if we adopt the smaller radius for the
radio photosphere, the masers still form in two distinct
rings: there is not a population of maser spots of
similar brightness extending from the outer ring inwards
towards the inner group, though the mapping software
used to generate Fig.~\ref{map_photos} suggests that
there may be.
 In fact, if we adopt a radio photospheric
radius of 1.39\,R$_{p}$, corresponding to the
2.03\,\micron \,pumping band, the spectra
and maps are almost identical to Model~1b, which features
only the outer ring. This intermediate radius is already
large enough to exclude all the very bright spots with
no observational counterparts.

An interesting feature of the outer ring (Fig.~\ref{map_photos})
is a `missing' sector in the (observational) South-West.
This arises purely from random selection of sites, but such
gaps do sometimes appear in the maser rings of real stars.

We adopt the larger radio photosphere from now
on, corresponding to the 8.13\micron \,pumping band. Therefore,
the extremely bright, inner, ring of masers is excluded.
However, the
fact that this inner ring can be produced raises an
interesting point about the origin of the maser sites.
If we assume that the motion of a typical maser spot is
outward, more than inward, then we expect new spots to
feed into the distribution near its inner edge, and
therefore close to the radio photosphere. How do these
new spots appear, and do they predominantly `switch on'
at a particular phase? The old model did not address this
question because it set an initial random distribution of
spots at model phase zero (when a new shock left the
stellar photosphere) and subsequently adjusted 
the positions of these spots with phase, following
the velocity field in the envelope. There was no
source of new spots. The new model may be able to
derive a phase for the appearance of a new ring once more
phase samples are available.

\subsection{Variation of dust envelopes}
\label{resdust}

\begin{figure}
\includegraphics[width=84mm]{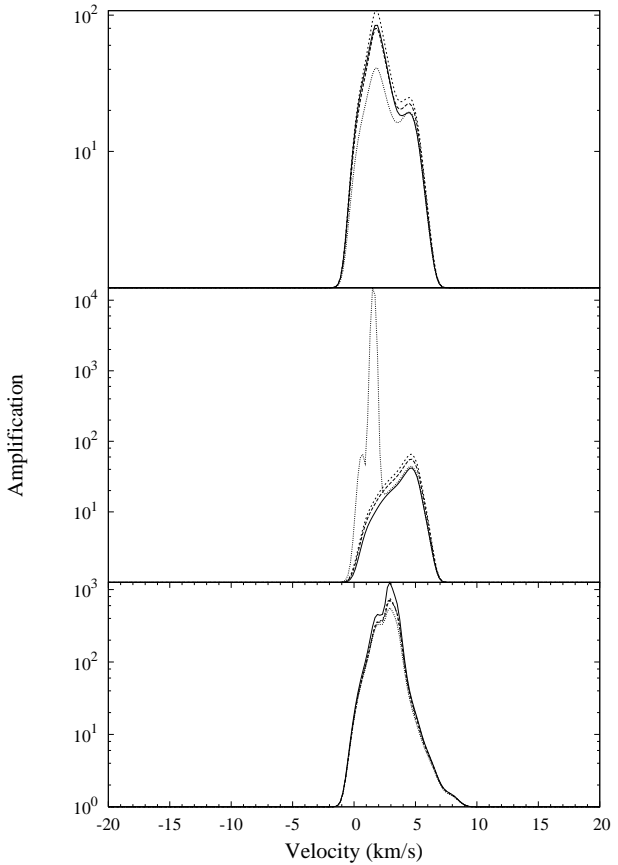}
\caption{Model spectra for the maser transitions $v=1, J=1-0$
(top panel), $v=1, J=2-1$ (middle) and $v=2, J=1-0$ (bottom
panel). Each panel has four versions of the spectrum
corresponding to: Model~1, dust epoch D, (solid line),
Model~2, dust epoch C, (long dashes),
Model~3, dust epoch B, (short dashes) and
Model~4, dust epoch A (dotted line). The hydrodynamic 
solution is M21n in all cases.
}
\label{spec_dust}
\end{figure}

We find that the effects of varying the dust epoch for the
single hydrodynamic solution, M21n, vary from subtle shifts to
profound change, depending on the transition in question. As
an example, we show in Fig.~\ref{spec_dust}, model spectra for three
commonly observed lines. Beginning with the top panel, the
43\,GHz $v=1, J=1-0$ line, we see that the spectra have very
similar shapes for all four epochs, with a fairly subtle change
in the amount of amplification introduced by varying the dust
radiation field. In this line, Model~3, with dust epoch~B, 
produces the brightest spectrum, with a peak amplification
factor of 108, followed by Model~1 (epoch~D dust) with 84,
Model~2 (epoch~C dust) with 80, and finally Model~4 (epoch~A dust)
with 41. Varying the dust radiation field therefore 
changes the peak maser amplification by a maximum factor
of just over 2 in this line. The $v=2, J=1-0$ line (bottom
panel in Fig.\ref{spec_dust}) also responds quite subtly
to changes in the dust envelope, with very similar shapes for
all four epochs. However the brightness order is different from
the analogous line in $v=1$: here Model~1 yields the strongest
line, though Model~4 is still the weakest.

In the 86\,GHz line ($v=1, J=2-1$) we see a more profound,
qualitative, effect of varying the parameters of the dust
envelope. Spectra for this transition appear in the middle
panel of Fig.~\ref{spec_dust}. Model~4, with
dust epoch~A, which yields the
weakest spectrum in both 43\,GHz lines, introduces a strong
additional peak 
on the blue side
of the spectrum at 86\,GHz, which is not present at the
other 3 epochs.

\begin{figure}
\includegraphics[width=84mm]{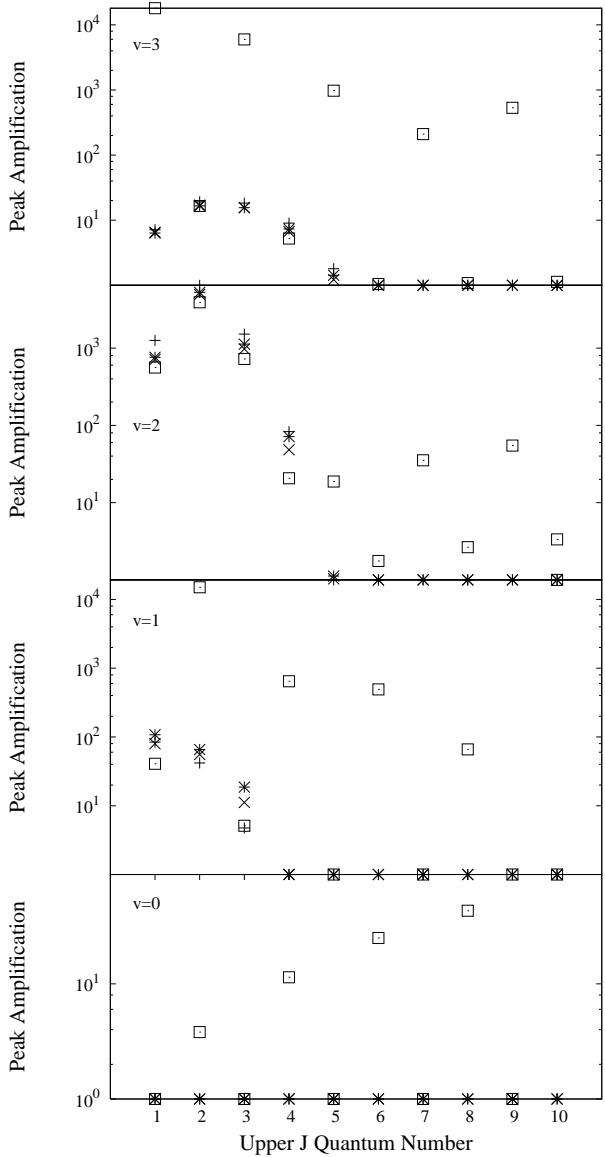}
\caption{Peak amplification factors found in the
spectra of the first ten rotational transitions
(upper $J$-quantum number on the x-axis) of the
first four vibrational states of SiO (panels as
marked with $v$ increasing from bottom to top. Values
for different dust epochs are marked with a different
symbol: $+$ for Model~1 (Epoch~D), $\times$ for
Model~2 (Epoch~C), $\ast$ for Model~3 (Epoch~B) and
$\Box$ for Model~4 (Epoch~A).
}
\label{peak_dust}
\end{figure}

A more general view of the effect of the dust radiation
field on the SiO masers is presented in Fig.\ref{peak_dust}.
Here we plot the peak amplification factor found in each
spectrum. Each panel of the figure is a different vibrational
state. Different symbols denote the dust epoch, and peak
values are marked against upper $J$-quantum number for 
transitions $J=1-0$ up to $J=10-9$.
The main point to be drawn from Fig.~\ref{peak_dust} is
that there is something special about Model~4, which uses
dust epoch~A, and is marked by the symbol $\Box$. Model~4
is the only one of the four to generate any masers in
$v=0$, where inversions are only generated in transitions
where the upper rotational level has even $J$. In the
other vibrational states, only Model~4 has masers for
upper $J>5$; at lower $J$, the peak amplification values
generated by the other three models are closely grouped,
while the value for Model~4 is often very different.
The middle panel of Fig.\ref{spec_dust} is an example
of this. The even-odd pattern produced by Model~4
in $v=0$ is also echoed through the other $v$-states:
masers with upper $J$ even are favoured in $v=0,1$, but
masers with upper $J$ odd are favoured in $v=2$ and $3$.
We cannot, at present, explain the peculiarity of Model~4
in detail, but we note that in Fig.\ref{newtdust}, 
the Epoch~A dust has the second-smallest inner dust shell
radius, but a significantly higher dust temperature
at its inner radius than the other three models. 

If the inner ring of masers is considered,
the effect of changing the dust parameters was 
much more marginal. The weaker, but `standard' outer ring
of spots are much more strongly affected individually,
and when grouped into a spectrum,
than those which were very intense.

\begin{figure}
\includegraphics[width=84mm]{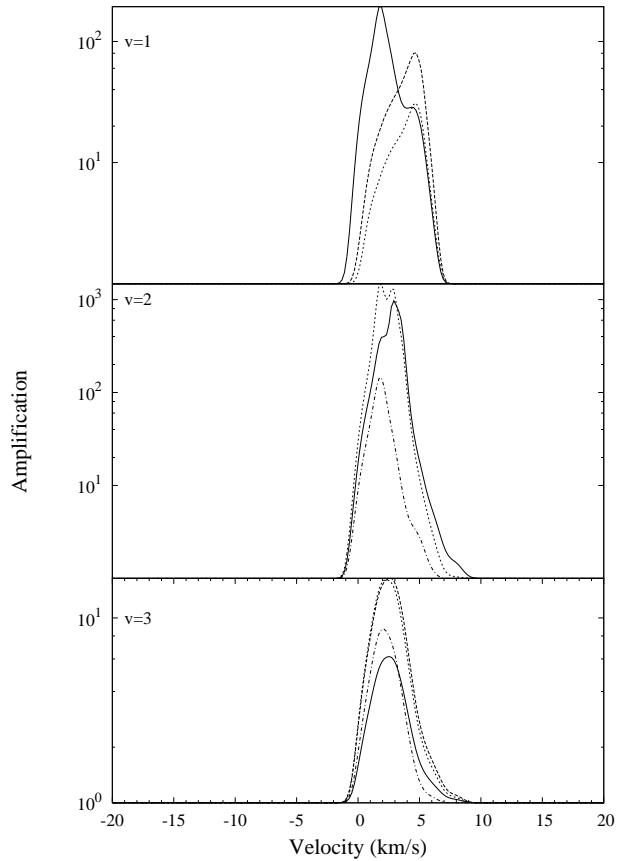}
\caption{Spectra of the masers found in a dust-free
model, based on the hydrodynamic solution M21n. Rotational
transitions have the following line styles:
$J=1-0$ (solid), $J=2-1$ (dashed), $J=3-2$ (dotted),
$J=4-3$ (chained).
}
\label{nodust}
\end{figure}

To test the 'absolute' effect of dust, an additional model was
run without a dust shell. In this model, pumping is entirely
due to the radiation from the star (in Model M21n) and kinetic
collisions. We show spectra for those lines in $v-1$, $v=2$
and $v=3$ which showed significant maser emission in
Fig.~\ref{nodust}. As elsewhere, the $v=2, J=2-1$ line
was excluded becasue of its interaction with a water transition.

When we compare the dust-free models to those with dust shells,
we find that the largest effect is in $v=1, J=1-0$. The shape
of the spectrum is similar, but if we compare the solid line
in the top panel of Fig.~\ref{nodust} to the lines in the
top panel of Fig.~\ref{spec_dust}, we see that the peak in the dust-free
case is brighter, by a factor of approximately 2, than the
brightest case with a dust shell. For the other lines in
Fig.~\ref{spec_dust} ($v=2, J=1-0$ and $v=1, J=2-1$) we again find
similar shapes (with the exception of dust epoch~A) and smaller
changes in maser amplification. Overall then, the presence of Al$_{2}$O$_{3}$
dust can have both positive and (mildly) negative effects on
maser pumping. We intend to investigate the effect of silicate dust,
if present, in future work.

\subsection{Phase Samples}
\label{resphase}

\begin{figure}
\includegraphics[width=84mm]{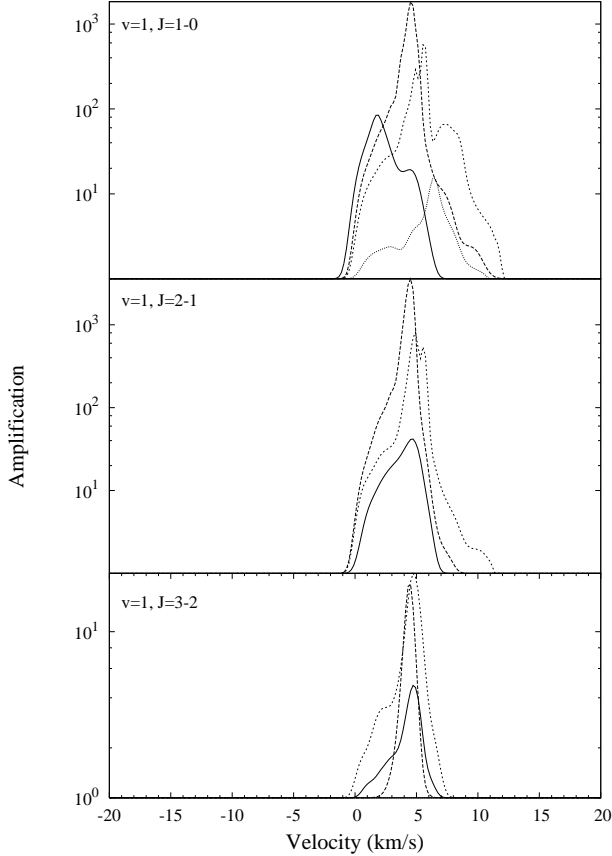}
\caption{Spectra of the bright masers in $v=1$ for the
phase samples in Table~\ref{phasetab}. Each panel is for
the transiton as marked. Spectra in each panel have the
following line styles: $\phi$=0.1, Model~1, (solid line),
$\phi$=0.25, Model~5, (long dashes),
$\phi$=0.30, Model~6, (short dashes) and
$\phi$=0.40, Model~7, (dotted line).
}
\label{v1spec}
\end{figure}

\begin{figure}
\includegraphics[width=84mm]{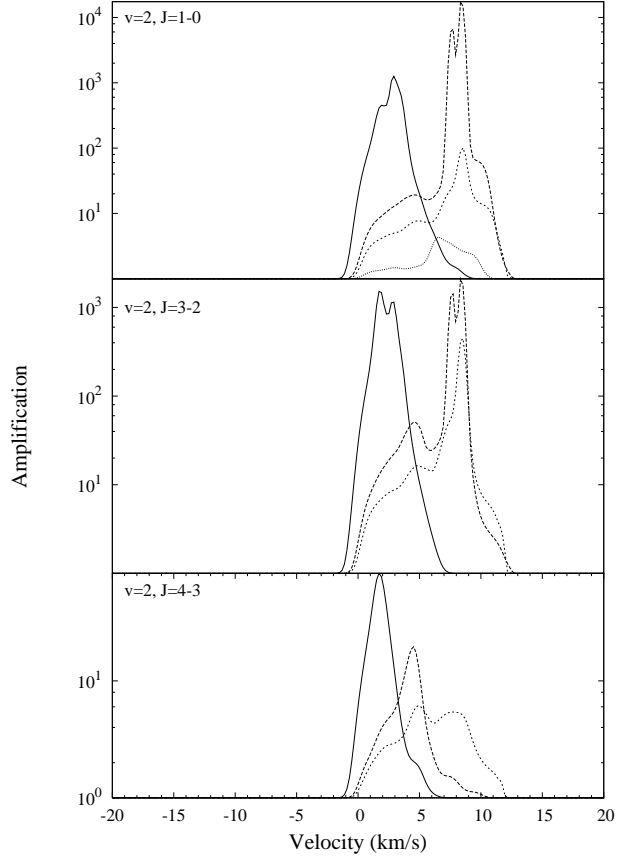}
\caption{As for Fig.~\ref{v1spec}, but for $v=2$. 
}
\label{v2spec}
\end{figure}

In the four phase samples listed in Table~\ref{phasetab}, no masers
were found in $v=0$. In $v=3$, weak masers were found in
rotational states up to $J=5-4$, with the greatest amplification
factor, of 19.2, at phase 0.1 (Model~1) in $v=3, J=2-1$.
All the $v=3$ masers followed a pattern of producing the
strongest maser emission at phase 0.1, with a monotonic decay to
phase 0.4. Only $J=1-0$ and $2-1$ masers survived to reach
phase 0.4. In Fig.~\ref{v1spec} we plot spectra of the three
masers from $v=1$ which achieve amplification factors of
greater than 20. These form in the lowest three rotational
transitions. Figure~\ref{v2spec} shows the spectra of bright
masers from $v=2$. Here, we ignore the $J=2-1$ line because of
its well-known interaction with a water transition, so the 
plot consists of the masers in the $J=1-0$, $J=3-2$ and
$J=4-3$ transitions.

The main points to note from Fig.~\ref{v1spec} are (i) the
$J=1-0$ line is the most stable: it is the only line which
still emits as a maser at phase 0.4; (ii) the most favourable
phase for masers is 0.25 (Model~8): both the $J=1-0$ and
$J=2-1$ lines are strongest at this phase, and in $J=3-2$ the maser
from the
next phase is only marginally stronger; (iii) in the $J=1-0$
line, there appears to be a systematic redward shift of the
spectral peak with phase, which does not seem to be the case
in the other two lines.

In Fig.~\ref{v2spec}, we find that,
as for $v=1$,  the $v=2, J=1-0$ line produces
the most stable spectrum, with the only
surviving emission at phase 0.4. The redward shift of the
spectral peak with phase is apparent for all three lines
shown in Fig.~\ref{v2spec}, with a very distinct shift between phases
0.1 and 0.25 in the $J=1-0$ and $J=3-2$ transitions. This
shift significantly changes the shape of the spectra,
while the peak moves to the red by $\sim$7\,km\,s$^{-1}$. The
brightest phase is 0.25 for the $J=1-0$ line, but 0.1 for
the other two.

\begin{figure}
\includegraphics[width=84mm]{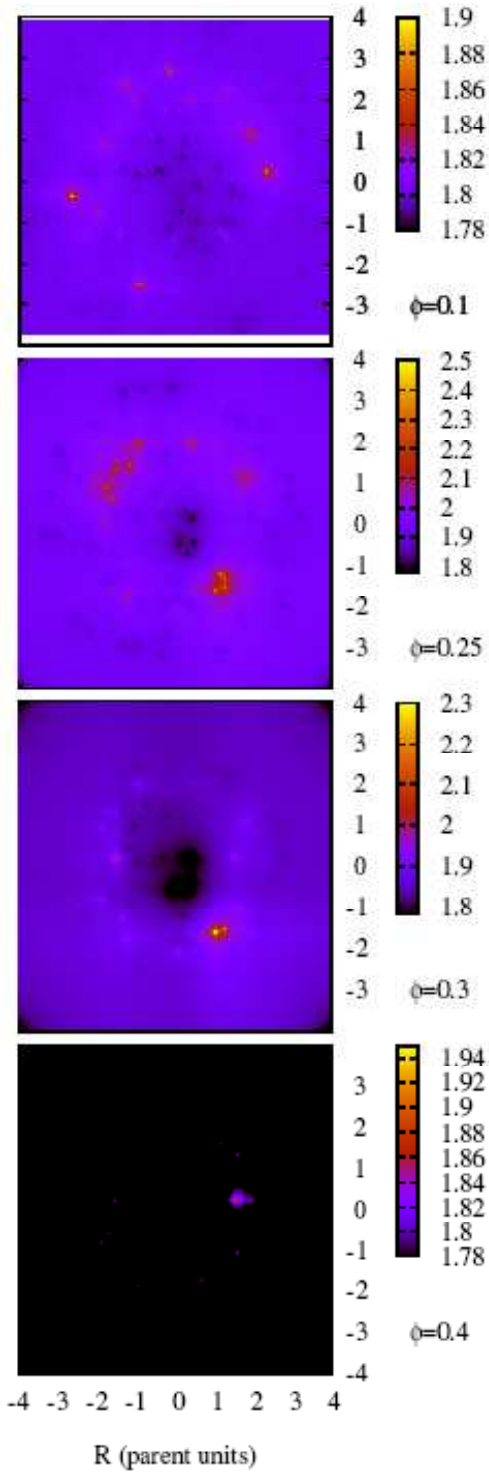}
\caption{Maps of the spot distribution for masers in
the $v=1, J=1-0$ transition for the four phase samples,
as marked. The colour scale is in the logarithm to the
base 10 of the amplification.
}
\label{mapv1J1_0}
\end{figure}

\begin{figure}
\includegraphics[width=84mm]{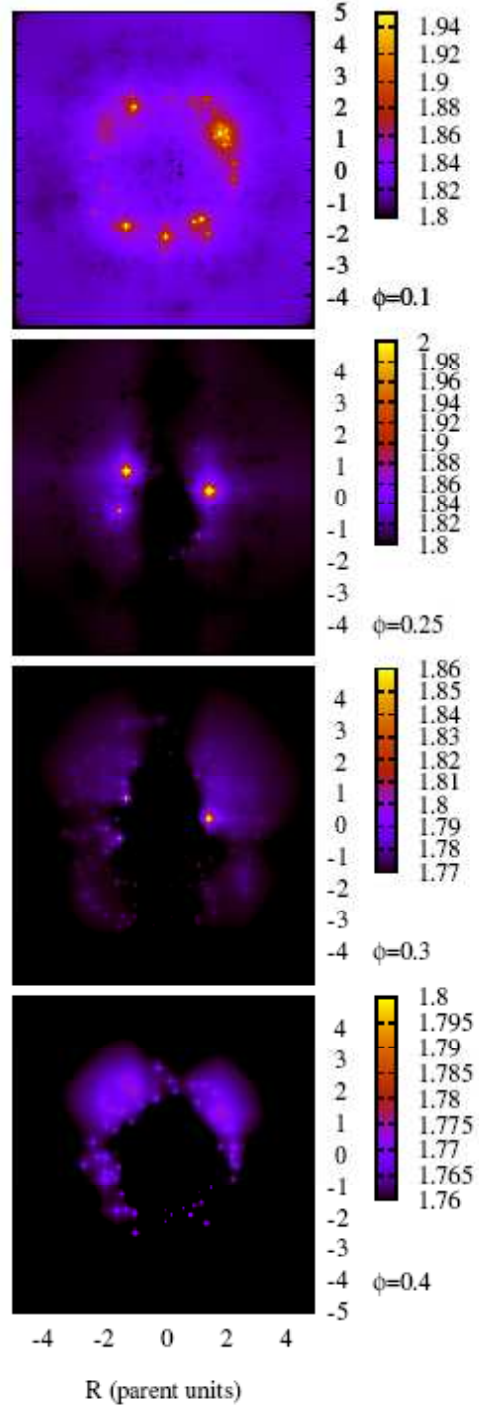}
\caption{As Fig.~\ref{mapv1J1_0}, but for the
$v=2, J=1-0$ transition.
}
\label{mapv2J1_0}
\end{figure}

\begin{figure}
\includegraphics[width=84mm]{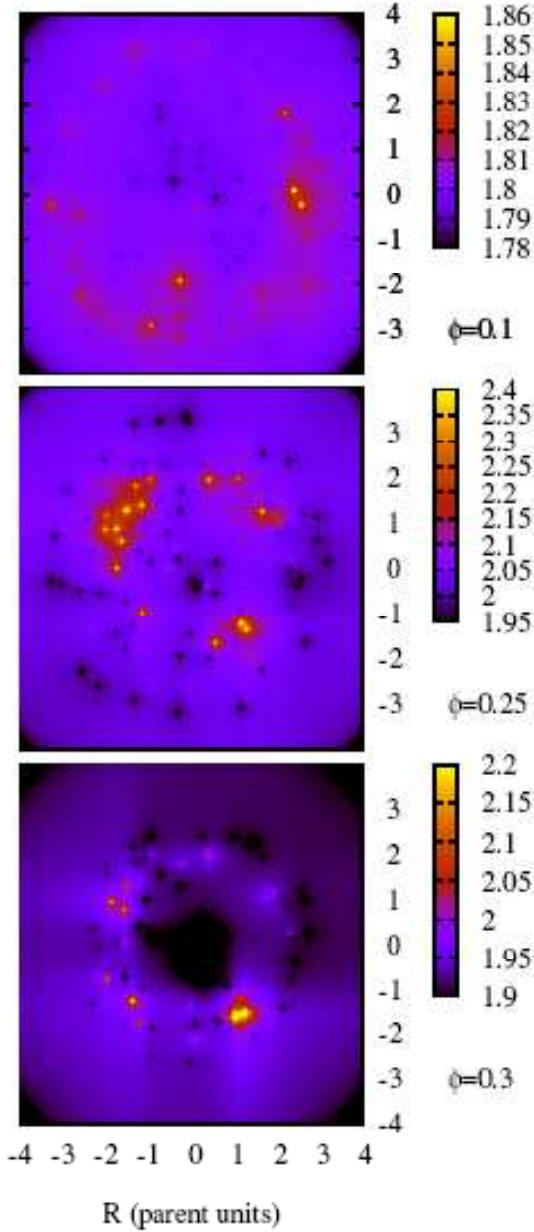}
\caption{As Fig.~\ref{mapv1J1_0}, but for the
$v=1, J=2-1$ transition. No map is shown for
phase 0.4 because the model produced no emission
in this line for this phase.
}
\label{mapv1J2_1}
\end{figure}

\begin{figure}
\includegraphics[width=84mm]{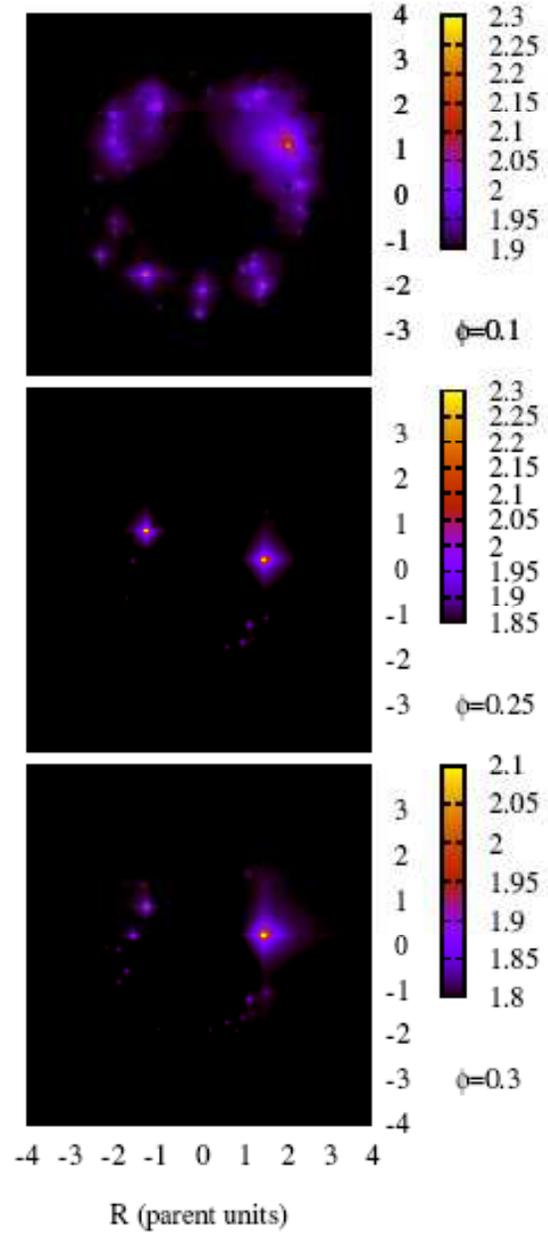}
\caption{As Fig.~\ref{mapv1J1_0}, but for the
$v=2, J=3-2$ transition. No map is shown for
phase 0.4 because the model produced no emission
in this line for this phase.
}
\label{mapv2J3_2}
\end{figure}

We now consider the spatial distribution of maser spots
as a function of stellar phase in
four transitions: $J=1-0$ in $v=1$ and $v=2$, $v=1, J=2-1$
and $v=2, J=3-2$. The $v=1, J=1-0$ maser shell is shown
in Fig.~\ref{mapv1J1_0}. The main change between phase 0.1
and the brightest phase (0.25) is  a decrease in
the ring radius, as judged by
the brightest spots. After phase 0.25, there is a fading of the
masers, with only a single obvious object remaining in the
maps for phase 0.4.
For the other 43\,GHz line, $v=2, J=1-0$, we show its
evolution with phase in Fig.~\ref{mapv2J1_0}. The behaviour
is broadly similar to that of its counterpart in $v=1$, but
it is clearer that, in the $v=2$ transition, something
catastrophic happens to the maser distribution between
phases 0.1 and 0.25. Although the masers are brighter in
the latter phase, the ring is smaller, indistinct, and
based on just a few very bright objects. These fade almost
to extinction by phase 0.4. It is noticeable that in the
map at phase 0.1, the region covered by weak maser emission
is larger than for $v=1$, so the $v=2$ maps are plotted out
to 5, rather than 4\,R$_{p}$. The apparent North-South line
of symmetry in the later phases in Fig.~\ref{mapv2J1_0} 
results only from the random spot distribution and the
mapping software, and does not result from any feature of
the hydrodynamic solutions or dust shell.
A comparison of Fig.~\ref{mapv1J1_0} and Fig.~\ref{mapv2J1_0}
phase by phase, suggests that, for the bright masers at least,
the $v=2$ spots occupy a smaller ring than those in $v=1$.
In Fig.~\ref{mapv1J2_1}, we show the maser shell of the 
86\,GHz transition, $v=1, J=2-1$. Only three phases are shown,
as there is no maser emission at phase 0.4. In this line, the
ring structure is distinct through the three phases shown,
but, as with the 43\,GHz lines, there is evidence of a
distinct change in structure, with a modestly strong, large
radius maser ring ($\sim$3\,R$_{p}$) replaced by an intense
ring of radius $<$2\,R$_{p}$.
The final set of maps we show, which is not yet observable
with a VLBI interferometer, is the $v=2, J=3-2$ transition
in Fig.~\ref{mapv2J3_2}. Like $v=1, J=2-1$, there is no
emission at phase 0.4. The first two phases have a very
similar peak amplification, as seen in the spectrum
in Fig.~\ref{v2spec}, but the spatial distribution of spots
has still suffered a similar change in distribution to
the other mapped transitions between phases 0.1 and 0.25.
This change is also reflected in the spectrum of this line.
The maser ring is arguably less rich than those from
lower rotational states.

Extended structure visible in the all the maps in this work
(Figs.~\ref{mapv1J1_0}-\ref{mapv2J3_2} 
and Fig.~\ref{map_photos}) should be viewed with
great caution. All the masers are actually pointlike
(see Section~\ref{prob_point}).
The {\sc GNUPLOT} software used to make the maps includes a
gridding algorithm which smooths the spot distribution into pixel
values on a regular grid. The colour palette is then applied
to these pixel values. However, there are typically several
hundred low amplification maser spots which, when gridded and
smoothed by the mapping software, yield the extended structure.
If our distribution of low amplification points generate
an equivalent flux to weak extended emission in real stars,
could they explain observed discrepancies between the integrated
flux from VLBI images and the flux from autocorrelation
spectra? If we simply divide our spots into two classes of
bright and weak, and assume that the masers in both classes
have the same sizes and beaming angles, then the ratio of fluxes
produced by the bright and weak objects is 
$N_{b}I_{b}/(N_{w}I_{w})$, where N is the number of objects,
subscripted with $b$ for bright, or $w$ for weak, and similarly
for brightnesses, or specific intensities, $I$. As an example,
we take the map for phase 0.1 in Fig.~\ref{mapv1J1_0}, and set the
dividing line at spots with amplification factors of 200.
We then have 6 bright and 256 weak spots. To obtain similar
fluxes from the weak and bright populations we then need an average
weak spot to be no fainter than 1/43 of the brightness of an
average bright spot. On this basis our distribution of weak
spots might possibly provide the flux missing from VLBI images. 

\section{Discussion}

The radio photosphere plays a vital role in reconciling the
new model with observations. By selecting a rather large
value of this parameter, a radius similar to 2\,R$_{p}$,
a group of extremely bright masers in an inner ring of
projected radius $\sim$1\,R$_{p}$ is excluded, leaving the
remaining masers in a ring which is consistent with both
observations and the old model. It is also useful to look
at what else is obscured by selecting the large radio photosphere.
In particular, the strong inner shock with
a velocity change of $>$20\,km\,s$^{-1}$, which is shown
in Fig.~\ref{newv} at a radius of 2.9$\times$10$^{13}$\,cm,
would not be visible in the radio. Our radio photosphere
at this phase (0.1) has a radius of 
2.2\,R$_{p}$, or 3.98$\times$10$^{13}$\,cm, so we would
only be able to see the outer, highly damped, shock in
Fig.~\ref{newv}, and the maximum observable velocity range in
the envelope at radio frequencies is 
only $\sim$7\,km\,s$^{-1}$. Velocities of this magnitude,
but not the 20\,km\,s$^{-1}$ of the inner shock, are
in agreement with observations 
\citep{reidmen07}, which measure the maximum velocity
from the proper motions of SiO masers, and link these spatially
to the radio photosphere with VLA continuum measurements. 
The damping of the shocks in the hydrodynamic
solutions used in the current work is therefore an important
improvement over the old model, in which the shocks were less
well damped, and propagated at higher speeds for a given
radius in the envelope.

The effects of dust can be both constructive and detrimental
to the maser pump. The most spectacular effect is that the
dust in Model~4 (epoch~A) has a qualitatively different effect on
maser emission when compared to the other three dust
epochs, used in conjunction with the same hydrodynamic
solution (M21n).
We investigate briefly the origin of this behaviour, for
which the $v=1, J=2-1$ line is a good example
(see Fig.~\ref{spec_dust}). On close examination, a single
bright model spot is responsible for much of the extra
peak in the spectrum of this line. This spot is positioned
at a radius (unprojected) of 3.59\,R$_{p}$, which is rather
large, but the ring radius for this line is also larger
than average (see discussion below and Fig.~\ref{mapv1J2_1}).
This radius places the bright spot well within the
epoch~A dust shell, very close to the 9th shell radius.
We compared the dust mean intensity at the spot radius
for all four epochs, and all the
SiO IR pumping bands, but there is nothing peculiar about
epoch~A in this respect. It is therefore not the direct
pumping power of the dust that makes epoch~A different.
A feature of the epoch~A shell which is different from
the other three epochs, with regard to the selected spot
position, is a combination of high dust temperature,
density  and 
considerable depth into the shell. Epoch~D has a higher
dust temperature (1070\,K versus 930\,K), but
the spot lies only between the 3rd and
4th radii. The spot lies
deeper inside the dust (near the 11th radius) in epoch~B,
but this shell has lower density and temperature (715\,K).
The most likely explanation for the particular behaviour
of masers with dust from epoch~A is therefore an optical
depth effect, with the dust modifying the 
stellar IR continuum in the pumping bands. We intend
to investigate the effects of dust further in future work.

The phase information in the present work is rather limited, so
we have not attempted to follow the evolution of maser rings
through the stellar cycle in a consistent way. We note also
that in the modelling of S~Ori, which used the the same
hydrodynamic solutions and dust shells, the fits to the observed
optical phases \citep{markus07}, and
see Table~\ref{epochtab}, have a sequence which does not
match that of the phases drawn directly from the theory.
In spite of this, and following the theory-based phases,
we find that the
maser peak for all lines is in the phase range
0.1 to 0.25. This is
consistent with the maser peak lagging the optical light
somewhat. Given the error of 0.1 in the absolute phases
from the model, this phase lag is consistent with typical
observed lags between SiO masers and the optical light curve.

A distinct feature of all the maps is a structural change
between phases 0.1 and 0.25, where the maser ring shows
a decrease in radius and, usually, an increase in
spectral output. This change in the structure of the map
is accompanied, in some lines, by a redward shift in the
spectral peak. There is also a shift from a large number
of maser spots of modest brightness, to a smaller number
of objects, a few of which are very intense. This behaviour
may have some underlying cause related to the envelope
dynamics, but as spot positions have been re-randomised, and
not evolved, between the two phases, we can only
suggest that this is an interesting feature, worthy of
further investigation, rather than a hard result.

The new model allows us to be considerably more
precise than the old model in studying the spatial relationship
of the maser rings to each other, and to other important
structures in the model, such as shocks, dust shells and
optically thick layers. We display this information in 
Table~\ref{bigtable}.
The first six columns of this table display model
parameters, which are independent of the maser model. The
remaining 6 columns show maser ring radii for the four transitions
mapped in Fig.~\ref{mapv1J1_0}-\ref{mapv2J3_2}. Of these
six columns, the first four are ring radii from the present
model, whilst the final two are from observations of
S~Ori, matched to theoretical phases as
in \citet{markus07}. All radii
given in Table~\ref{bigtable} are relative to the IR
photosphere (at 1.04\,\micron ) for the particular phase for
easy comparison with \citet{markus07}, Table~7 and Fig.~12.

Without even considering any maser ring radii, we can see
that the 8.13\micron \, band radius (Column~6) is closely related
to the position of the innermost shock (Column~3). The
radii are the same to 0.1\,R$_{1.04}$ fo three epochs, but
markedly different for epoch D (phase 0.1). This finding
strongly suggests that the whole way of thinking about
shocks propagating vast distances from some almost-fixed stellar
photosphere, which caused many problems for the old model
is simply outdated. For a significant portion of the cycle
at least, the 8.13\micron \,radius
is tied to a shock, so that
the delay between irradiation and shock impact would
be 25\,d, or less. Consequently,
many arguments 
which eliminate some element of collisional pumping on the
grounds of long shock propagation times, relative to a
distant, detached, stellar photosphere, are not relevant to resolving
the pumping debate. The physical changes in the maser shells
between phases 0.1 and 0.25 may be caused by the shift from
a state where the 8.13\,\micron \,radius and inner shock
are detached, to one where they are closely coupled.
\begin{table*}
 \centering
 \begin{minipage}{140mm}
  \caption{Key radii for maser shells, and their association with
shocks and optically thick band radii. All radii are written in terms of the
IR photospheric value, $R_{1.04}$, at the appropriate
phase. Columns are as follows: 1 - the hydrodynamic solution and
dust epoch (the same as the observing epoch in \citet{markus07});
2 - the stellar phase from theory (optical scale); 3 - radius of
the inner shock; 4 - radius of the outer shock (not apparent
at phase 0.4); 5 - inner radius of the dust shell;
6 - optically thick band radius in the $\Delta v=1$ SiO pumping
band; 7 to 10 - maser ring radii (in sky projection)
with sample standard deviations
for the transitions marked from the present model;
11 and 12 - observed maser ring radii from \citet{markus07} for
transitions as marked. In columns 3 and 4, we have listed the
radii for the same shocks. At phase 0.4, the original outer
shock has damped out, so we have left the
entry for $R_{s2}$ blank at this phase. A new shock has
formed at phase 0.4, with a radius of 1.0\,R$_{1.04}$, inside
the original inner shock, listed as $R_{s1}$.
}
\label{bigtable}
  \begin{tabular}{@{}llrrrrrrrrrr@{}}
  \hline
  \multicolumn{6}{c}{Model Parameters} &
  \multicolumn{6}{c}{Maser Ring Radii (photospheric units)}\\
Model+dust & $\phi$ & R$_{s1}$ & R$_{s2}$ & R$_{in}$ & R$_{8.13}$ &
R$_{43.1}$&R$_{42.8}$&R$_{v=1,J=2-1}$&R$_{v=2,J=3-2}$&
R$_{43.1}^{obs}$&R$_{42.8}^{obs}$ \\
 \hline
M21n+D & 0.1 & 1.3 & 2.5 & 2.4 & 1.8 & 
   1.9$\pm$0.4 & 1.9$\pm$0.4 & 2.4$\pm$0.7 & 2.0$\pm$0.4 &
      -        &     -       \\
M22+A  & 0.25& 1.6 & 2.6 & 1.8 & 1.6 &
   1.9$\pm$0.3 & 1.8$\pm$0.3 & 1.8$\pm$0.5 & 1.8$\pm$0.3 &
   2.2$\pm$0.3 & 2.1$\pm$0.2 \\
M23n+C & 0.3 & 1.7 & 2.8 & 2.2 & 1.7 &
   1.9$\pm$0.2 & 1.8$\pm$0.2 & 1.9$\pm$0.2 & 1.9$\pm$0.2 &
   2.1$\pm$0.3 & 1.9$\pm$0.2 \\
M24n+B & 0.4 & 2.0 & -   & 2.0 & 2.0 &
   2.1$\pm$0.5 & 2.4$\pm$0.4 &    -        &    -        &
   2.4$\pm$0.3 & 2.3$\pm$0.4 \\
\hline
\end{tabular}
\end{minipage}
\end{table*}

The outer shock appears to be well outside the maser zone,
except at phase 0.1, where it may be associated with the
large $v=1, J=2-1$ ring (see Column~9). At most phases,
the inner shock, the 8.13\,\micron \,radius and the
maser ring radii of all the lines are within 0.3\,R$_{1.04}$
of each other. This association does not involve the
inner shock at phase 0.1. 
The inner edge of the Al$_{2}$O$_{3}$ dust, R$_{in}$ (Column~5) is less well
associated with masers, inner shock and 8.13\,\micron \,radius.
It is close to the maser ring radius at phase 0.25 and 0.4, but
not at phases 0.1 and 0.3.

The maser rings themselves are closely associated for almost
all transitions and phases. The exception is the large ring
at phase 0.1 for the $v=1, J=2-1$ transition. The ring radius
for this transition is not peculiar at the other phases. In
the first three phases, the mean ring radius of the 
$v=2, J=1-0$ masers is the same, or
slightly smaller than that for the
analogous transition in $v=1$. This trend is swapped for phase
0.4, but this phase has a large standard deviation, due to the
small number of surviving spots, and a rather poorly defined
ring. At this phase, close to optical minimum, the masers
from both $J=1-0$ (43\,GHz) transitions have their largest
mean ring radii in both observations and theory.
The model is therefore consistent with previous models
and with VLBI observations which show that the $v=2, J=1-0$
ring is usually slightly smaller. The large ring for
the $v=1, J=2-1$ masers at phase 0.1 is not in agreement
with the old model, but is more in line with VLBI
observations \citep{soria07}.
The modelled ring radii at 43\,GHz
are also smaller than their observational counterparts, for the
three phases where observational data exist. However, the
ring sizes are consistent, given the standard deviations
attached to these ring radii.

Maser spectra in the 43\,GHz ($J=1-0$) lines have velocity
extents of $\sim$10\,km\,s$^{-1}$, a value which is consistent
with the observational spectra of these lines towards
S~Ori \citep{markus07}. The LSR velocity of S~Ori is not
very well known, so we have not computed Doppler biases
for our spectra, though these model spectra are certainly
concentrated to the red side of the velocity range.

Perhaps the largest drawback of the current work, in terms of
matching VLBA observations of S~Ori, is in the variation of
spectral intensity. In the three observed phases, the
ratio of the spectral maxima for the brightest and dimmest
phases is about 1.5 for the $v=1, J=1-0$ line, and
about 2 for $v=2, J=1-0$. By contrast, the fading of the
model masers towards phase 0.4 gives analogous ratios of approximately
70 and 1000. However, as has been noted previously, our
models are not evolved in phase, but use repositioned spots,
so the numbers could be very different if the masers followed
the gas motions. There are also only a small number of
bright spots, so these ratios are very vulnerable to
statistical fluctuations.

\section{Conclusions}

The new combined maser and hydrodynamic solutions 
produce maser emission in rings of typical size
2.2 IR photospheric radii. The ring structures are similar
in many respects to those produced by the old model.
An analysis of ring radii confirms the usual observational
result, and the prediction of the old model, that the
$v=2$ ring is smaller than the $v=1$ ring at 43\,GHz,
whilst a large ring is possible at some phases in the
$v=1, J=2-1$ transition at 86\,GHz. The overall range
of maser ring radii, from 1.8-2.4 IR photospheric radii,
for the 43\,GHz transitions in
$v=1$ and $v=2$ is also consistent
with VLBI experiments where maser rings and photospheric
radii have been compared at the same phase
\citep{bob05,fedele05,markus07}.

If a radio photosphere of approximately 2 IR photospheric
radii is used, observable shock velocities are
in agreement with actual radio continuum observations,
as high velocity shocks, such as the inner
shock at phase 0.1, are hidden.
A radio photosphere of this size also prevents the appearance
of extremely bright masers very close to the
IR photosphere.

Dust can both suppress and enhance maser emission, and
there is definitely a radiative component to the pumping
scheme. However, the close association of shock and
8.13\,micron \,radii suggests that collisional
and radiative pumping are closely associated spatially,
and therefore temporally.

Further work is required to consider the effects of
dust in detail, and to produce a fully phase-sampled
model of one or more complete stellar cycles.

\subsection*{ACKNOWLEDGMENTS}

MDG acknowledges STFC (formerly PPARC) for financial support under
 the 2005-2010 
rolling grant, number PP/C000250/1.
Computations
were carried out at the HiPerSPACE Computing Centre, UCL, which is
funded by the UK STFC. The authors would like to thank Lee Anne
Willson for detailed comments on the manuscript.

\end{document}